# Electron-Phonon Interaction in Ternary Rare-Earth Copper Antimonides LaCuSb$_2$ and La(Cu$_{0.8}$Ag$_{0.2}$)Sb$_2$ probed by Yanson Point-Contact Spectroscopy

N.V.Gamayunova, D.L.Bashlakov, O.E.Kvitnitskaya,
A.V.Terekhov, Yu.G.Naidyuk

B.Verkin Institute for Low Temperature Physics and
Engineering of NAS of Ukraine,
47 Nauky Ave., Kharkiv, 61103, Ukraine
gamayunova@ilt.kharkov.ua

Z.Bukowski, M.Babij

W. Trzebiatowski Institute of Low Temperature and
Structural Research of Polish Academy of Sciences,
Ul. Okólna 2, 50-422 Wrocław, Poland

**Investigation of the electron-phonon interaction (EPI) in LaCuSb$_2$ and La(Cu$_{0.8}$Ag$_{0.2}$)Sb$_2$ compounds by Yanson point-contact spectroscopy (PCS) has been carried out. Point-contact spectra display a pronounced broad maximum in the range of 10÷20 mV caused by EPI. Variation of the position of this maximum is likely connected with anisotropic phonon spectrum in these layered compounds. The absence of phonon features after the main maximum allows the assessment of the Debye energy of about 40 meV. The EPI constant for the LaCuSb$_2$ compound was estimated to be $\lambda$=0.2±0.03. A zero-bias minimum in differential resistance for the latter compound is observed for some point contacts, which vanishes at about 6 K, pointing to the formation of superconducting phase under point contact, while superconducting critical temperature of the bulk sample is only 1K.**

*Keywords—point-contact spectroscopy; ternary rare-earth antimonides; LaCuSb$_2$; La(Cu$_{0.8}$Ag$_{0.2}$)Sb$_2$*

## I. INTRODUCTION

In the frame of the research attention to iron-based superconductors (IBS), it might be of interest to compare these materials with another similar superconducting family. Besides the recently discovered 112-type of IBS [1,2], there is another group of the superconducting compounds with the same structural type – the ternary rare-earth antimonides RETSb$_2$ (RE is a rare-earth metal, T is a transition metal) – with the same structural type. The latter contains a rare-earth element and antimony instead of arsenic but does not contain iron [3]. This family including the superconducting systems LaNiSb$_2$, LaCuSb$_2$, LaPdSb$_2$ and others was investigated thoroughly concerning their magnetic and transport properties by Muro et al. [3], Myers [4], and thermoelectric properties by Ohta [5].

After the discovery of IBS, the family of the ternary rare-earth antimonides attracted strong attention again due to richness their electronic properties and relative simplicity their crystal structure. They crystallize in a tetragonal ZrCuSi$_2$ - type (P4/nmm) structure [6], which looks very similar to that of the 112-type of IBS [1,2]. The study of the superconducting features of the ternary rare-earth antimonides is of great interest in view of the possibility to compare such two superconducting families.

Here we present the study of the LaCuSb$_2$ and La(Cu$_{0.8}$Ag$_{0.2}$)Sb$_2$ compounds carried out by using Yanson PCS, a powerful method to study the fundamental processes of conduction electron scattering and their interaction with other quasiparticles and excitations in solids [7]. Yanson PCS could be helpful in reveal of details of the EPI as well as, to resolve other quasiparticle interactions, what can help to shed light on the nature of superconducting state in the materials under study.

## II. EXPERIMENTAL DETAILS

High-quality single crystals of LaCuAgSb$_2$ and La(Cu$_{0.8}$Ag$_{0.2}$)Sb$_2$ were grown using the self-flux method as described in ref. [8].

The samples are layered single crystals with lateral dimension of 2x1x1 mm$^2$. They show the typical metallic behavior of the resistivity and become superconducting below $T_c \approx 1$ K and $T_c \approx 0.5$ K for LaCuSb$_2$ and La(Cu$_{0.8}$Ag$_{0.2}$)Sb$_2$ respectively. The residual resistance ratio of LaCuSb$_2$ sample was about 8, what is slightly higher compared to that of a sample from the one of the first publication [9], which indicates a good quality of the crystal.

These materials have been investigated in the normal state by Yanson PCS method to obtain the information about the EPI function. The observed nonlinearities of the current-voltage characteristics (and their derivatives) for point contacts between two metals contain information about the phonon spectrum. If the characteristic size (diameter) of the point contact is less than the electron mean free path, the electrons can pass through the constriction ballistically (without scattering). The electrons with excess energy $eV$ ($e$ is the electron charge, $V$ is the applied voltage) from one side of the contact scatter inelastically creating the phonons in the investigated sample. This scattering will appear by the additional negative contribution to the current flowing through the point contact. Such deviations allow obtaining the energy resolved information about the interaction processes of the electrons with the elementary excitations (phonons, magnons, etc.) in the investigated material. The applied voltage defines the energy of the corresponding interaction process. This property allows determining the spectral EPI function for many elementary metals and various conducting compounds [7].

We have measured the first and second derivatives of the current-voltage characteristics (CVC) for the homo- and heterocontacts on the base of $LaCuSb_2$ and $La(Cu_{0.8}Ag_{0.2})Sb_2$ compounds. In the case of using heterocontacts with noble metals, we expected more probable realization of spectral regime due to the low residual resistance of these metals and correspondingly the large mean free path of electrons.

The point contacts were formed using the conventional needle-anvil method by touching a cleaved single crystal surface of a sample with a sharpened thin Cu or Ag wire in the case of heterocontacts or between two clean cleaved surfaces of the samples in the case of homocontacts [7]. The precision two-coordinate mechanism enables to change the pressing force of the electrodes and to probe through the sample surface.

The resistance $R$ of the point contacts was in the range from several Ohms to tens Ohms, what corresponded to the diameter $d$ of the contacts from several units to hundreds of nanometers, according to well know formula for the ballistic regime [7]:

$$d = \sqrt{\frac{16\rho l}{3\pi R}} \qquad (1)$$

Here, we used $\rho l = p_F/e^2 n = 5.3 \times 10^{-12}$ $\Omega cm^2$ as for copper, since the calculated Fermi energy ($\approx 7.1$ eV) of $LaCuSb_2$ in ref. [6] is almost equal to the Fermi energy of copper,

The point contacts were formed in a liquid helium cryostat at T=4.2K, which is higher than the superconducting transition temperature $T_c$ in the investigated samples ($T_c \approx 1$ K for $LaCuSb_2$ and $T_c \approx 0.5$ K for $La(Cu_{0.8}Ag_{0.2})Sb_2$). PCS of EPI is also possible in the superconducting state, but the superconducting features due to the effects of Andreev reflection or the critical current (or magnetic field of current) effects are much more intense than the contribution of EPI in the point-contact spectrum and overlap the latter. Therefore, the point-contact spectra of the EPI were usually measured in the normal state at a temperature higher than the critical temperature.

The first derivatives of CVC (or differential resistance) $R_D=dV/dI(V)$ for such point contacts and the second derivatives of CVC (or point-contact spectra) $d^2V/dI^2(V)$ were measured by detecting of the first and second harmonics of the modulating signal proportional to the corresponding derivatives of CVC with a lock-in amplifier.

According to PCS theory [7], the second derivative of the CVC curve $d^2V/dI^2=RdR/dV$ is directly proportional to the EPI spectral function $\alpha^2_{PC}F(\varepsilon=eV)$:

$$\frac{1}{R}\frac{dR}{dV}(eV) = \frac{8ed}{3\hbar v_F}\alpha^2_{PC}F(\varepsilon)\big|_{\varepsilon=eV}, \qquad (2)$$

where $d$ is the diameter of the point contact, and $v_F$ is the Fermi velocity.

III. RESULTS AND DISCUSSION

The differential resistance $R_D=dV/dI(V)$ for such point contacts (Fig. 1, 2 insets) shows the metallic behavior with a low value of the parameter $\Delta R/R$, that is one of the criteria of the spectral (ballistic or diffusive) regime [7] of the current flow throw the contact.

The point-contact spectra of $LaCuSb_2$ and $La(Cu_{0.8}Ag_{0.2})Sb_2$ display one pronounced broad maximum in the range of $10 \div 20$ mV for the both samples (see Fig. 1, 2, 3 main panels). The appearance of a pronounced maximum at the finite values of bias voltage relatively to the total intensity of the spectrum followed by the output of the featureless background is also a distinguishing evidence of the spectral regime.

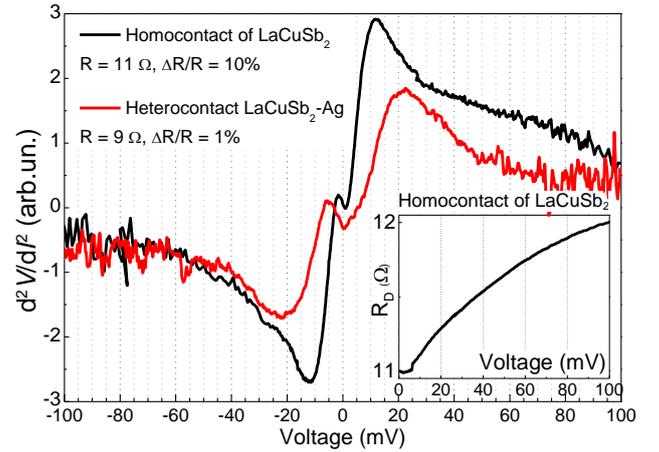

Fig.1. Point-contact spectra for two contacts of $LaCuSb_2$ measured at 4.2K. Inset: differential resistance $R_D=dV/dI$ for point contact with $R=11\Omega$.

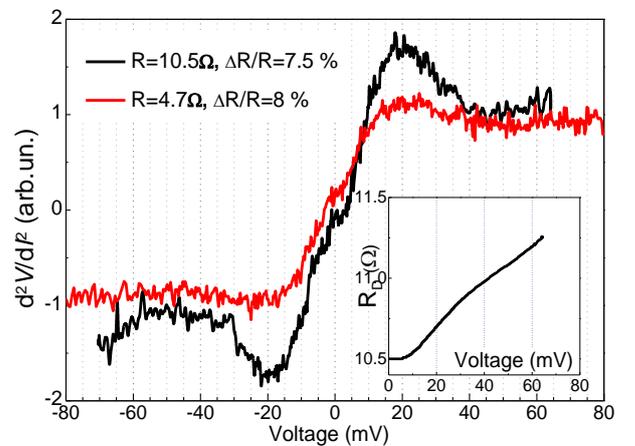

Fig.2. Point-contact spectra for two homocontacts of $La(Cu_{0.8}Ag_{0.2})Sb_2$ measured at 4.2K. Inset: differential resistance $R_D=dV/dI$ for point contact with $R=10.5\Omega$.

Firstly we had to ensure that the obtained spectral information comes directly from the sample and it is not connected with contribution from the copper or silver counter-electrodes in the case of using heterocontacts, because the transverse phonons of copper and silver, may contribute to the point contact spectra in the range between 15 and 20 meV in Cu [7] and at 11–13 meV in Ag [7]. For this we have carried the measurements with the both copper and silver needles and also prepared the homocontacts for both samples. The spectra of the homocontacts demonstrate the same features like in the case of the heterocontacts, testifying negligible contribution of noble metals to the point-contact spectra.

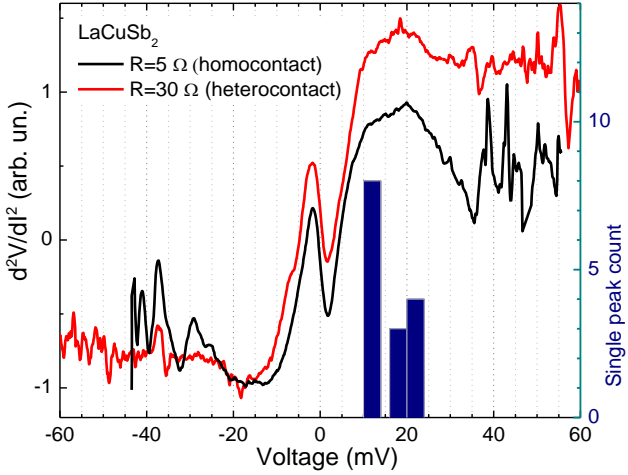

Fig. 3. Two rare point-contact spectra with the "double" feature at ≈10 and ≈20 mV. Statistical data set demonstrating the position of the maxima for different point-contact spectra for LaCuSb$_2$ are shown at the bottom.

Let us discuss the nature of the observed features. The varying position of the maximum in the range 10÷20 mV (see statistical data on Fig. 3) can be due to the strong anisotropy in these layered materials. Such spectral features located at the characteristic energies of the phonon frequencies in metals, and these peaks can be associated with EPI in the investigated materials. The measurements of the point contacts in magnetic field (Fig. 4) do not change the shape and position of the main features, what confirms their phonon nature.

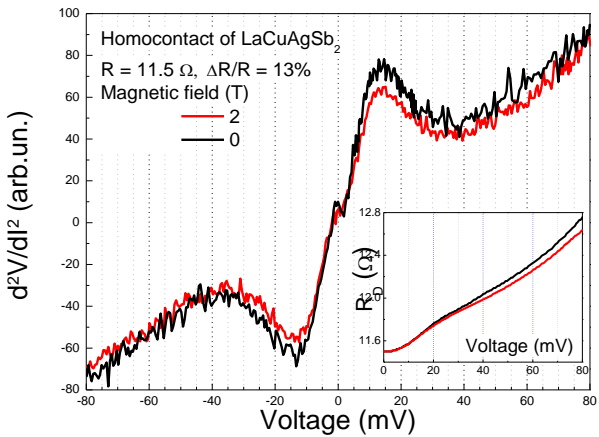

Fig.4. Point-contact spectra for homocontacts of La(Cu$_{0.8}$Ag$_{0.2}$)Sb$_2$ measured at 4.2K in magnetic field (2 T) and without magnetic field. Inset: differential resistance $R_D=dV/dI$ for both point contacts.

In general, we have not found principal difference in the point-contact spectra between LaCuSb$_2$ and La(Cu$_{0.8}$Ag$_{0.2}$)Sb$_2$, because of the variation of the spectra for the same compound. That is the anisotropy of the spectra in this case is more stronger than the difference in the characteristic phonon frequencies.

Featureless spectra after the main maximum allow to evaluate the Debye energy of about 40 meV, what is in contrast with the low Debye temperature of 151 K (≈ 13meV), estimated from the specific heat measurements in ref. [3]. It is interesting that the point-contact spectra of another ternary compound LaCu$_2$Si$_2$ display the main maximum in similar region of 10÷20 mV [10] and Debye energy about 40meV, although this system has different crystal structure.

For LaCuSb$_2$ compound, we evaluated EPI constant λ

$$\lambda = 2\int_0^\infty \frac{\alpha^2 F}{\varepsilon} d\varepsilon, \quad (3)$$

using $\alpha^2 F(\varepsilon=eV)$ from (2) and the value of the Fermi velocity $v_F$ in (2) again like in cooper. The obtained maximal value of EPI constant is 0.2±0.03. This value corresponds to the lower limit of $\lambda$, because the calculated spectra indicate rather the diffusive regime with broad phonon maxima [11] instead of the ballistic regime with sharper phonon features.

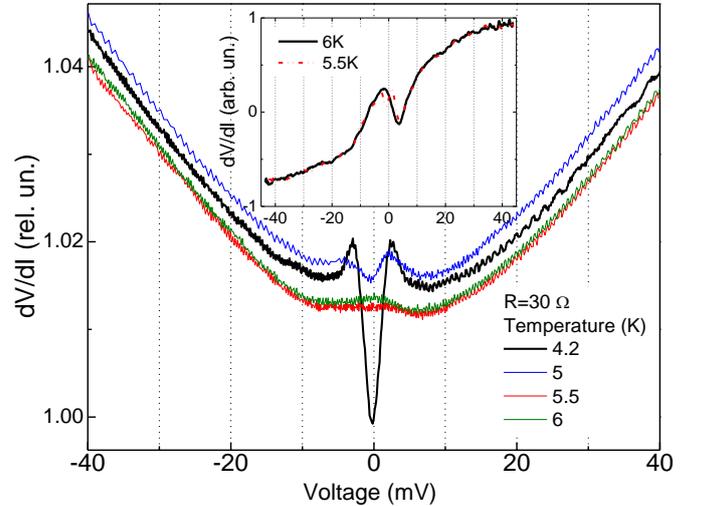

Fig.5. Differential resistance $R_D=dV/dI$ for LaCuSb$_2$ heterocontact with resistance of 30 Ω measured at different temperatures. Inset shows second derivative for the same contact measured at two temperatures.

For several point contacts we observed differential resistance $dV/dI$ with a zero-bias sharp minimum (Fig. 5). The minimum disappears around 6 K and it is likely connected with superconductivity in the core of the point contact. It is known that the face-centered cubic form of La has a transition temperature of 6.3 K, whereas that of the hexagonal close-packed form has $T_c \approx 4.9$ K [12]. The second derivative, in this case, does not display spectral features like phonon maximum in the range above 10mV (see Fig. 5 inset), that is the material in the point contact is disordered. Whether this minimum is related to the superconducting La cluster or the superconducting temperature of LaCuSb$_2$ can increase in point contact due to pressure effects, it is necessary to investigate further.

Another features in the point-contact spectrum are, so-called, zero-bias N-shaped anomaly, often seen at $V$=0 (see, e.g., Fig. 3). This features in $d^2V/dI^2$ corresponds to the shallow zero-bias maximum in the differential resistance $dV/dI$. The most probable reason for this anomaly is the Kondo effect, that is, scattering on magnetic impurities [13] or presence of two-level systems in the case of strong disorder [14]. In both cases it indicates not perfect crystal structure under the point contact at the surface.

In conclusion, we investigated EPI in LaCuSb$_2$ and La(Cu$_{0.8}$Ag$_{0.2}$)Sb$_2$ compounds by Yanson PCS. The point-contact spectra display the pronounced broad maximum in the range of 10÷20 mV caused by EPI. Variation of positions of this maximum is likely connected with anisotropic phonon spectrum in these layered compounds. Featureless spectra after the main maximum allow estimating Debye energy of about 40 meV. The calculation of EPI constant $\lambda$ for LaCuSb$_2$ gave the low limit of 0.2±0.03 for its value. For some point contacts a zero-bias minimum in differential resistance was observed, which vanishes at about 6K, testifying formation of superconducting phase under the contact.

ACKNOWLEDGMENT

N.V.G., D.L.B., O.E.K., A.V.T., Yu.G.N. acknowledge support the National Academy of Sciences of Ukraine under project Ф4-19. N.V.G. thanks for the support from ILT&SR within the scholarship of 2016 granted by ILT&SR for ILTPE employees.